# C Based Hardware Design for Wireless Applications

Andres Takach
andres_takach@mentor.com
*Mentor Graphics*

Bryan Bowyer
bryan_bowyer@mentor.com
*Mentor Graphics*

Thomas Bollaert
thomas_bollaert@mentor.com
*Mentor Graphics*

## Abstract

*The algorithms used in wireless applications are increasingly more sophisticated and consequently more challenging to implement in hardware. Traditional design flows require developing the micro architecture, coding the RTL, and verifying the generated RTL against the original functional C or MATLAB specification. This paper describes a C-based design flow that is well suited for the hardware implementation of DSP algorithms commonly found in wireless applications. The C design flow relies on guided synthesis to generate the RTL directly from the untimed C algorithm.*

*The specifics of the C-based design flow are described using a simple DSP filtering algorithm consisting of a forward adaptive equalizer, a 64-QAM slicer and an adaptive decision feedback equalizer. The example illustrates some of the capabilities and advantages offered by this flow.*

## 1. Introduction

The growth in wireless communication has been fueled by the application of modern DSP algorithms that enable the adaptation to varying communication channel characteristics and efficient usage of channel bandwidth. Wireless communication are present in consumer applications such as cell phones and local area networks and span a wide range of data rate and channel requirements.

The computational nature of DSP algorithms used in wireless applications are well suited to such a design methodology. Figure 1 shows how the proposed methodology fits into existing flows. The main difference compared to a traditional design flow is the replacement of the manual transformation of the C into RTL with an automated synthesis flow, where the designer guides synthesis to generate the micro architecture to meet the desired performance/area goals. The C synthesis product used in this paper is *Catapult C*. Synthesis generates the RTL with detailed knowledge of the delay of each component to eliminate the guess work that is otherwise unavoidable when the micro architecture and RTL are generated manually. The advantages of an automated C-based synthesis flow are reflected both in significantly reduced design times as well as higher quality of designs, because a variety of micro architecture can be rapidly explored.

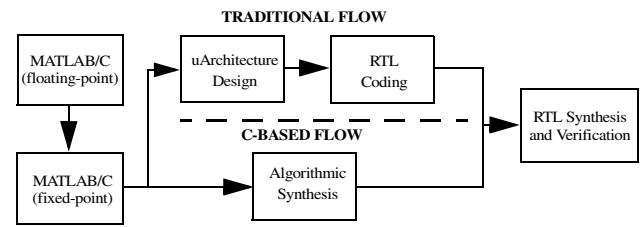

**Figure 1:** C-Based Design Flow

The typical design flow for implementing DSP algorithms starts with writing the algorithm at a functional level using languages such as MATLAB, C or a combination of the two languages (C here is used to refer to both C and C++). Due to its faster execution speed, C is typically preferred over MATLAB for modules being implemented in hardware. These modules are often the most computationally intensive, making them the most demanding to simulate.

In many cases, algorithms are initially written using floating point arithmetic and are then *numerically* refined to use finite-precision arithmetic. The refined algorithms are bit-accurate specifications.

In general, the effort of moving from a bit-accurate algorithmic functional description to a synthesizable algorithmic C implementation model is relatively low. No complexities of timing or concurrency or target technology are encoded in these models. To verify the refined C models are simulated against the original MATLAB/C models.

For each C algorithmic model the designer explores different architectures by directing how data will move in an out of the block (interface synthesis), mapping arrays to memories, and deciding how much parallelism is required to meet the throughput/latency goals. In traditional flows, defining the architecture and generating the RTL from the C model is done manually, a process that may required several months to complete. In a C based synthesis flow, the architecture definition and the generation of RTL is often accomplished in a matter of days to weeks.

Simulation is used for individual blocks or to debug designs, but in general is too slow to perform functional verification of the system. The generated RTL can be used to obtain an FPGA prototype of the RTL that is then used for functional verification.

If the FPGA prototype of the RTL is not sufficiently fast enough to allow "at-speed" emulation of the ASIC block, a new FPGA design can be rapidly generated that does match the speed of the ASIC design. Using such a strategy, the designer can perform functional field testing to validate the algorithm(s) used or validate the RTL for other blocks of the design or other components of the system.

The paper is organized as follows. Section 2 presents an overview of algorithmic C synthesis and the transformations that are essential in that process. Section 3 provides an introduction to important issues that need to be taken into account when coding the C algorithm for synthesis. Section 4 outlines the example DSP algorithm and provides much of the C behavior for the design including a templatized class for complex numbers based on fixed-point arithmetic. Section takes the algorithm through architectural exploration to generate RTL with certain performance goals and presents the result for a number of architectures. The paper closes with conclusions that can be drawn from our work.

## 2. Algorithmic C synthesis

The synthesis flow starts with a purely algorithmic C specification. The absence of explicit timing or concurrency makes algorithmic specifications far more compact and implementation independent than traditional RTL or "behavioral" specifications written in languages such as VHDL, Verilog or SystemC.

The hardware architecture is obtained by applying architectural directives during synthesis. Architectural directives provide a mechanism to specify high level decisions on how the design communicates with the outside world, how data is stored and how parallelism is exploited to obtain the desired performance. For example, in the algorithmic C, the input data is passed when the function is called, and the "outputs" of the function are available to the caller when the function returns. During synthesis, the designer specifies through *interface synthesis* directives how the data is transferred from and to the design.

The main architectural transformations that are used to generate a hardware architecture from an algorithmic C specification include:
- interface synthesis
- variable/array mapping
- loop pipelining
- loop unrolling
- scheduling.

The transformations are described in more detail in the following sections.

### 2.1 Interface synthesis

Interface synthesis converts the way the C function communicates with the outside world. There are a number of architectural transformations that take place that allow writing the C specification in such a way that does not require embedding features of the desired architecture in the source. The transformations include:
- An optional start/done handshake protocol is added to the design
- The individual C function arguments are mapped to a variety of resources such as memories, buses, FIFOs, handshaked registers etc.
- The data transfer bitwidth for any of the arguments is specified
- Arrays accesses over an index may be converted into accesses over time. For example, an array "uint10 x[1024]" may generate a port of width 10 bits that is read over time throughout the execution of the algorithm to access the different x[i]. The environment then *streams* in the data in the required order, most commonly in increasing or decreasing index order.

### 2.2 Variable/array mapping

Interface or local variables or arrays may be mapped to memories or may be split into registers. Smaller arrays are typically mapped to registers while larger arrays are typically mapped to memories. The required read and write bandwidth of the memory depends on the algorithm and the performance requirements on the design.

### 2.3 Loop pipelining

Loop pipelining provides a way to increase the throughput of a loop (or decreasing its overall latency) by initiating the $(i+1)^{th}$ iteration of the loop before the $i^{th}$ iteration has completed. Overlapping the execution of subsequent iterations of a loop exploits parallelism across loop iterations. The number of cycles between iterations of the loop is called the initiation interval.

In many cases loop pipelining may improve the resource utilization, thus increasing the performance/area metric of the design.

## 2.4 Loop unrolling

Partial or full unrolling of a loop exposes parallelism that exist across subsequent iterations of the loop. In some cases, partial unrolling may also be used in a coordinated way with memory mapping and interface synthesis to increase the effective bandwidth for data transfer. For example, unrolling may expose the possibility of accessing of even and odd elements of an array as one word when it is mapped to memory.

## 2.5 Scheduling

Scheduling is the focal point of architectural exploration. It transforms the sequential specification into a architecture with a well defined cycle-by-cycle behavior. It takes into account required synthesis directives such as the clock period and the target technologies. In addition, it takes into account cycle and resource constraints that are either explicitly provided by the user or implied by interface synthesis directives, variable/array mapping directives and loop pipelining/unrolling directives. Scheduling selects among combinational, sequential and pipelined components that implement the operations in the algorithm.

## 3. Coding C for synthesis

In general, it is best to start from simpler and more compact code. It is also important to understand the hardware implications of different C/C++ constructs. For instance, memory allocation/deallocation (malloc, free, new, delete) is not supported. In this Section we highlight some general guidelines for coding C for synthesis.

The function for the design is designated as the top-level design using a pragma. Arguments of the function are used by interface synthesis to generate the appropriate ports. Arguments such as "int *x", or "int x[10]" or "int &x" are treated as *in*, *out*, or *inout* depending on whether the object pointed at is only read, only written or, both read and written. Non pointer arguments such as "int x" are treated as input ports and read at the start of the execution of the algorithm.

## 3.1 Bit-accurate datatypes

Hardware designers are accustomed to bit-accurate datatypes in hardware design languages such as VHDL and Verilog. The C language defines a number of integral types and floating-point types. While floating-point types are useful for simulating DSP algorithms, they are impractical for most hardware implementations. *Catapult C* supports the native C integer types as well as bit-accurate integer and fixed-point types. Native C integer types provide signed and unsigned datatypes of bitwidths 1 (bool), 8 (char), 16 (short), 32 (int), 64 (long long). Bit masking, shifting or C bitfields may be used to model and synthesize bitwidths from 1 to 64 with the native C integer types.

*Catapult C* supports several C++ templatized classes that encapsulate bit-accurate behavior for integer and fixed-point types:

- <u>Limited precision integer types</u>: SystemC's sc_int/ sc_uint [4] and *Catapult C*'s mc_bitvector. They are limited to 64-bits and are relatively fast to simulate.
- <u>Arbitrary-length integer types</u>: it supports arbitrary bitwidths making the semantics far cleaner as operations return full (integer) precision:
  - SystemC's sc_bigint/sc_biguint: much slower simulation than limited-precision integer types.
  - *Catapult C*'s mc_int [3]: 3x to 100x faster simulation than SystemC integer types. It also has cleaner synthesis semantics than SystemC's integer types.
- <u>Arbitrary-length fixed-point type</u>: sc_fixed/sc_ufixed. They model fixed point datatypes with a variety of quantization and overflow modes. Slower than sc_bigint/sc_biguint.

## 3.2 Coding bitwidths

Specifying the actual bitwidth is mostly required for interface variables. It is often possible to take an algorithm written with C built-in integral types and only constrain the bitwidths of some variables at selected places:

- at the interface: if a 17 bit signed integer is required, we need to use int17 (part of mc_int) or sc_int<17> or sc_bigint<17> (SystemC datatypes).
- force a certain precision where synthesis is not able to reduce the bitwidth thorough its dataflow analysis. For example: the casting to int17 in the expression "a = (int17) (a + b*c)" would allow the reduction of variable "a" from a 32 bit integer (assume "a" was declared int) to a 17 bit integer.

Automatic bit reduction minimizes the changes required to generate optimized RTL from an existing C algorithm. It also makes writing of parameterized designs easier and less error prone. A typical example is a loop with a constant bound that depends on template parameters or constants that come from a "#define" statement in the source. The example C code is shown in Figure 2 where the minimum bitwidth required for loop variable i depends on the template parameter N.

```
template<int N>
int f(int *x) {
   int a = 0;
   for(int i=0; i < N; i++){
      a += x[i];
   }
   return a;
}
```

**Figure 2:** Example where minimum bitwidth of i depends on template constant

It is important to roughly know what bitwidths are expected as synthesis may not always be able to perform the level of analysis of an experienced designer. Such situations are normally found by examining the bill-of-materials report, the critical-path report, or by careful examination of the schedule (Gantt chart) as operations with long delays stand out. Cross probing from the reports or the Gantt chart to the source provides a quick mechanism to find out which expression or variable was not reduced in bitwidth.

Integer are often used to implement fixed-point arithmetic by using explicit calls to functions that perform rounding and saturation. Using integer to model fixed-point requires careful attention to the position of the implicit fixed-point.

SystemC provides signed and unsigned fixed-point datatypes that provide all the arithmetic operators and rounding/saturation functionality. For example, and sc_fixed<8,3,SC_RND,SC_SAT> is a fixed point number of the form bbb.bbbbb (8 bits of width, 3 bits integer) and with quantization mode set to SC_RND and overflow mode set to SC_SAT. The default modes (as when the sc_fixed is declared as sc_fixed<8,3>) is SC_TRN (simple truncation) and SC_WRAP (no saturation).

## 4. Case study: 64-QAM decoder

Multilevel quadrature amplitude modulation (M-QAM) [1][2] is used in many digital communication applications such as modems to efficiently and reliably use the communication channel. The digital signal is coded in levels, both in-phase and quadrature components leading to a two-dimensional constellation. The number of constellation points represent the number of values that may be coded in a symbol. For our design, we have an 8x8 constellation resulting in 64 codes or six bits of information per symbol.

In most communication applications the channel properties are not known in advance and in fact vary with time. Adaptive filtering is used to equalize the channel properties and to correct for *inter symbol interference* (ISI). The typical arrangement is shown in Figure 3. The feed-forward equalizer (FFE) performs channel equalization, while the decision feedback equalizer (DFE) corrects for the ISI from the N previously received symbols. The error signal is used to update the coefficients of the adaptive filters. We used the sign-LMS (least mean squared) adaptive algorithm for our example.

The input signal is complex, representing the in-phase and quadrature components of the received signal. All filters operate with complex data and have complex coefficients.

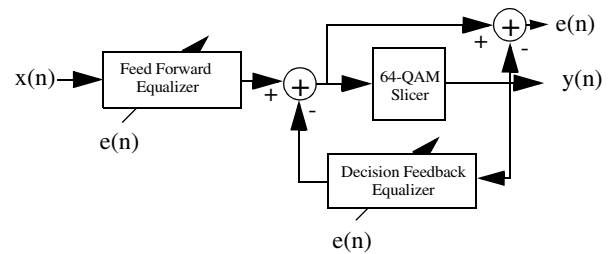

**Figure 3:** Block diagram for Equalized QAM decoder

For simplicity, we have not implemented details of how the training sequence is generated or blind adaptation is performed. Also we have not considered timing recovery within our design.

The FFE is T/2 spaced (taking two new inputs every symbol period) and has eight taps. The DFE is T spaced and has 16 taps. The filter coefficients are complex and have 10 bits of precision.

### 4.1 Coding of the algorithm

The algorithm is written based on the block diagram shown in Figure 3. The first version of such an algorithm is typically written using floating point arithmetic. Once the algorithm is validated, the algorithm can be refined to use fixed-point datatypes. The coding of the algorithm using fixed-point datatypes is shown in Figure 4. The algorithm is written so that the various bitwidths can easily be set by changing the definition of a few constants. Written the algorithm in a parametrized manner simplifies precision exploration and facilitates future reuse of the algorithm.

The process of minimizing the precision of the arithmetic requires knowledge of communication theory and DSP hardware design [5] as well as empirical (simulation based) analysis and validation. For the QAM-decoder, quantization of each computation is a source of noise that contributes to the overall mean-squared error at the input of the QAM slicer. The maximum mean-squared error that can be tolerated depends on the required bit-error rate. In addition to quantization noise, errors due to overflow need to be taken into account. The focus here, however, is in the hardware generation from a numerically refined algorithm.

The algorithm was written using a datatype that encapsulates complex arithmetic. The use of such a datatype helps to make the algorithm specification more compact and easier to understand but otherwise does not affect the synthesis process presented here.

The complex datatype is named sc_complex as it uses the sc_fixed datatype to represent the real and imaginary components. The sc_complex class was written by the authors (not part of the SystemC datatypes) and is not shown here due to space constraints. It is templatized much the same way as the sc_fixed datatype is.

Every call to the function qam_decoder takes two new inputs and computes a new 6-bit output. The arrays that store the tap values and the coefficients are declared *static* so that the values are preserved between calls to the function.

The various loop were given labels (part of C/C++) to facilitate the discussion on what loops will be transformed when doing architectural exploration. The design has a total of six loops (two for filter computation, two for adaptation, two for shifting the tap values). The "#define"s for the constants FFE_W, DFE_W, FFE_C_W, DFE_C_W are not shown but are all set to 10.

The loop nfe (dfe is similar) is a simple loop that implements an FIR filter. The coefficients of the nfe filter are updated in the loop ffe_adapt using the current error "e" and the last nffe "x" samples (inputs to the nfe filter). The ffe_shift loop shifts the "x" samples two places to make room for the two new input samples that are provided in the next invocation of qam_decoder.

## 5. Results

The target design is a 5 MBaud (30 Mbps) design using a 100MHz clock using an specific ASIC technology. In addition, we want to explore other designs to show how C synthesis may be used to rapidly generate various architectures.

The QAM_decoder function generates one output (symbol) for every invocation of the function. In order to meet the throughput requirements, the algorithm should take 20 or fewer cycles to compute a new output. An inspection of the algorithm reveals that a sequential execution of the six loops alone would take 8+16+8+16+3+15 = 66 cycles, roughly three times slower

```
#pragma design top
void qam_decoder( sc_complex<X_W,0> x_in[2],
   uint6 *data) {
   const int nffe = 8;
   const int ndfe = 16;
   const sc_fixed<FFE_C_W,0> mu_ffe =
      (sc_fixed<FFE_W+2,2>)1 >> 8;;  // pow(2, -8)
   const sc_fixed<DFE_C_W,0> mu_dfe =
      (sc_fixed<DFE_W+2,2>)1 >> 8;;  // pow(2, -8)

   // coeffs for forward and decision equalizers
   static sc_complex<FFE_C_W,0> ffe_c[nffe];
   static sc_complex<DFE_C_W,0> dfe_c[ndfe];
   static sc_complex<X_W,0> x[nffe];
   static sc_complex<4,0> SV[ndfe];
   static sc_complex<FFE_W,0> e;

   x[0] = x_in[0]; x[1] = x_in[1];

   sc_complex<FFE_W+1,1> yffe = 0;
   nfe: for(int k=0; k < nffe; k++)
      yffe += x[k]*ffe_c[k];// forward equalizer
   sc_complex<DFE_W+1,1> ydfe = 0;
   dfe: for(int k=0; k < ndfe; k++)
      ydfe += SV[k]*dfe_c[k];// decision feedback equalizer

   sc_complex<FFE_W+1,1> y = yffe - ydfe;// equalizer output

 // 64-QAM slicer
   sc_fixed<4,0> offset = 0;  offset[0] = 1;  // pow(2,-4)
   sc_fixed<3,0> r = (sc_fixed<FFE_W,0,
      SC_RND_ZERO,SC_SAT>) (y.r() - offset);
   sc_fixed<3,0> i = (sc_fixed<FFE_W,0,
      SC_RND_ZERO,SC_SAT>) (y.i() - offset);
   SV[0] = sc_complex<3,0>(r,i) +
      sc_complex<4,0>(offset, offset);
   sc_complex<FFE_W,0> e = SV[0] - y;
   sc_fixed<6,6> data_f = r*64 + i*8;
   *data = data_f.to_int();

   // Sign-LMS Adaptation for FFE and DFE
   ffe_adapt: for (int k=0; k < nffe; k++)
        ffe_c[k] += mu_ffe*e*x[k].sign_conj();
   dfe_adapt: for (int k=0; k < ndfe; k++)
        dfe_c[k] -= mu_dfe*e*SV[k].sign_conj();

   ffe_shift: for(int k=nffe-4; k >= 0; k-=2) {
      x[k+3] = x[k+1];
      x[k+2] = x[k];
   }
   dfe_shift: for(int k=ndfe-2; k >= 0; k--)
      SV[k+1] = SV[k];
}
```

**Figure 4:** C++ code for QAM decoder algorithm

than our goal.

In order to get a quick overview of the design, we run synthesis with the default architectural constraints (loop merging enabled, no loop unrolling/pipelining). The result is a design with a latency of 35 cycles. It turns out that synthesis merged loops dfe into loop ffe and merged loops dfe_adapt, ffe_shift and dfe_shift into loop ffe_adapt. Each loop takes 16 cycles to complete resulting in a latency of 3+16+16 cycles (three cycles for behavior between loops). With loop merging disabled, the design would take a latency of 3+8+16+8+16+3+15 = 69 cycles.

The latency of the design is improved by exposing additional parallelism. We found that loops of 8 and 16 iterations that were merged could be reduced in latency by partially unrolling the loops with 16 iterations (dfe, dfe_adapt, dfe_shift) to take eight iterations each (UNROLL=2). The resulting latency is 3+8+8 = 19 cycles, meets the performance design goals.

It is easy to exploit additional parallelism to further improve the latency of the design. For instance, setting unroll=2 for dfe and unroll=4 for dfe_adapt and dfe_shift reduces the latency to 3+8+4 = 15 (6.67 MBaud or 40 Mbps). Table 1 summarizes exploration results for our design. The area number is normalized with respect to the second design. The first six columns show the loops in the algorithm and the architectural options that were selected where "M" denotes that loop merging is enabled and "U" denotes the loop unrolling number.

**Table 1:** Comparison of architectures generated from C synthesis

| Architectural Loop Constraints | | | | | | Latency (ns) | Data Rate (Mbps) | Area |
|---|---|---|---|---|---|---|---|---|
| ffe | dfe | ffe_adapt | dfe_adapt | ffe_shift | dfe_shift | | | |
| M | M | M | M | M | M | 350 | 17.1 | 1.17 |
| none | none | none | none | none | none | 690 | 8.6 | 1.00 |
| M | M, U=2 | M | M, U=2 | M | M, U=2 | 190 | 31.5 | 1.61 |
| M | M, U=2 | M, U=2 | M, U=4 | M | M, U=4 | 150 | 40 | 1.88 |

The architectural exploration above was performed in a matter of minutes without changing the source description. We have also successfully targeted FPGA technologies. It is often possible to prototype the design at-speed with an FPGA. Such a prototype can be used to validate the algorithm itself (including the quantization decisions) as well as facilitate the hardware emulation of the whole system.

An architectural choice not shown in the above exploration is loop pipelining. Although it is possible to perform loop pipelining with this approach, for this algorithm and the given performance goals, loop pipelining does not provide as much benefit as loop unrolling. The main reason is that the loop body is simple enough that each iteration of the loop can be executed in a single cycle. Loop unrolling and loop pipelining can often be used in conjunction to effectively exploit parallelism in an algorithm.

## 6. Conclusions

A C-based design methodology was presented and used to generate RTL for a DSP algorithm commonly used in wireless designs. We showed the feasibility of the design methodology and provided a general overview of architectural C synthesis capabilities, C coding for synthesis and architectural exploration based on the *Catapult C* synthesis tool. A single C source was used to rapidly generate a set of ASIC designs with a wide range of data rate numbers. In addition, the C-based design methodology enables FPGA prototyping flows to facilitate verification of the generated RTL as well as real-time functional verification of the algorithm.

## 7. Acknowledgements

Authors would like to acknowledge and thank the work of the Catapult-C high-level synthesis team at Mentor Graphics without which this paper would not have been possible.